\documentclass[12pt]{article}
\pdfoutput=1
\usepackage{xcolor}
\usepackage{geometry}                
\usepackage[parfill]{parskip}    
\usepackage[pagewise]{lineno}
\usepackage{rotating}
\usepackage{amssymb}
\usepackage{amsmath}
\usepackage{rotating}
\usepackage{multicol}
\usepackage{multirow}
\usepackage{apacite}
\usepackage{graphicx}
\usepackage{array}
\usepackage{styles}
\usepackage{caption}

\usepackage{float}
\usepackage{subcaption}
\usepackage{tabularx}
\usepackage{longtable}
\usepackage[ruled,vlined,linesnumbered]{algorithm2e}

\usepackage[frozencache,cachedir=minted-cache]{minted}
\setminted{fontsize=\small}
\usepackage{listings}
\definecolor{backgroundCol}{rgb}{.97, .97, .97}
\usepackage{etoolbox}
\makeatletter
\patchcmd{\minted@colorbg}{\medskip}{}{}{}
\patchcmd{\endminted@colorbg}{\medskip}{}{}{}
\makeatother

\usepackage{etoolbox}
\makeatletter
\patchcmd{\@verbatim}
  {\verbatim@font}
  {\verbatim@font\small}
  {}{}
\makeatother

\newcolumntype{Y}{>{\centering\arraybackslash}X}
\usepackage{mathtools}

\usepackage{hyperref}
\makeatletter
\renewcommand\hyper@natlinkbreak[2]{#1}
\makeatother
\hypersetup{
    colorlinks=false, 
    linktoc=all,     
    linkcolor=blue,  
        linkbordercolor=blue, 
    pdfborder = {0 0 1}
}

\usepackage{psfrag,epsf}
\usepackage{enumerate}

\usepackage{secdot}
\usepackage[title,toc,titletoc,page]{appendix}

\begin{document}
\title{ddtlcm: An R package for overcoming weak separation in Bayesian latent class analysis via tree-regularization}

\author{Mengbing Li\textsuperscript{1,}\footnote{Corresponding: Mengbing Li (mengbing@umich.edu); Zhenke Wu (zhenkewu@umich.edu).} , 
Bolin Wu\textsuperscript{2}, 
Briana Stephenson\textsuperscript{3}, 
Zhenke Wu\textsuperscript{1,$\ast$}\\
\textsuperscript{1}Department of Biostatistics, University of Michigan\\
\textsuperscript{2}Department of Computer Science, University of Michigan \\
\textsuperscript{3}Department of Biostatistics, Harvard University 
}
\maketitle


\section*{Summary}

Latent class model (LCM) is a model-based clustering tool frequently used by social, behavioral, and medical researchers to cluster sampled individuals based on categorical responses. Traditional applications of LCMs often focus on scenarios where a set of unobserved classes are well-defined and easily distinguishable. However, in numerous real-world applications, these classes are weakly separated and difficult to distinguish. For example, nutritional epidemiologists often encounter weak class separation when deriving dietary patterns from dietary intake assessment \citep{li2023tree}. Based on the number of diet components queried and the heterogeneity of the study population, LCM-derived dietary patterns can exhibit strong similarities, or weak separation, resulting in numerical and inferential instabilities that challenge scientific interpretation. This issue is exacerbated in small-sized study populations. 

To address these issues, we have developed an R package \texttt{ddtlcm} that empowers LCMs to account for weak class separation.
This package implements a tree-regularized Bayesian LCM that leverages statistical strength between latent classes to make better estimates using limited data. With a tree-structured prior distribution over class profiles, classes that share proximity to one another in the tree are shrunk towards ancestral classes \textit{a priori}, with the degree of shrinkage varying across pre-specified item groups defined thematically with clinical significance. The \texttt{ddtlcm} package takes data on multivariate binary responses over items in pre-specified major groups, and generates statistics and visualizations based on the inferred tree structures and LCM parameters. Overall, \texttt{ddtlcm} provides tools designed to enhance the robustness and interpretability of LCMs in the presence of weak class separation, particularly useful for small sample sizes.

\section*{Statement of Need}

A number of packages are capable of fitting LCMs in R. For example, \texttt{poLCA} \citep{linzer2011PoLCAPackage} is a fully featured package that fits LCMs and latent class regression on polytomous outcomes. \texttt{BayesLCA} \citep{whiteBayesLCAPackageBayesian2014} is designed for Bayesian LCMs for binary outcomes. \texttt{randomLCA} \citep{beath2017randomlca} provides tools to perform LCMs with individual-specific random effects. These packages focus on LCMs where the class profiles are well-separated. Directly applying these packages to data that suffer from weak separation may result in large standard deviations of the class profiles, tendency to merge similar classes, and inaccurate individual class membership assignments, phenomena highly prevalent in small-sized study populations.

The package \texttt{ddtlcm} implements the tree-regularized LCM proposed in \cite{li2023tree}, a general framework to facilitate the sharing of information between classes to make better estimates of parameters using limited data. The model addresses weak separation for small sample sizes by (1) sharing statistical strength between classes guided by an unknown tree, and (2) accounting for varying degrees of shrinkage across major item groups. The proposed model uses a Dirichlet diffusion tree (DDT) process \citep{neal2003density} as a fully probabilistic device to specify a prior distribution for the class profiles on the leaves of an unknown tree (hence termed ``DDT-LCM"). Classes positioned closer on the tree exhibit more profile similarities. The degrees of separation by major item groups are modeled by item-group-specific diffusion variances.

\section*{Usage}

In the following, we use an example list of model parameters, named ``parameter\_diet", that comes with the package to demonstrate the utility of \texttt{ddtlcm} in deriving weakly separated latent class profiles. We start with demonstrating how a simulated dataset is generated. We next apply the primary model fitting function to the simulated dataset. Finally we summarize the fitted model and visualize the result.

\subsubsection*{Data Loading}
The ``parameter\_diet" contains model parameters obtained from applying DDT-LCM to dietary assessment data described in \cite{li2023tree}. We use it as a semi-synthetic data-generating mechanism to mimic the weak separation issue in the real world. Specifically, the data set includes a tree named ``tree\_phylo" (class "phylo"), a list of $J = 78$ food item labels, and a list of $G = 7$ pre-defined major food groups to which the food items belong. The food groups are dairy, fat, fruit, grain, meat, sugar, and vegetables.

\begin{minted}[breaklines,breaksymbolleft=\quad,bgcolor=backgroundCol]{R}
install.packages("ddtlcm")
library(ddtlcm)
# load the data
data(parameter_diet)
# unlist the elements into variables in the global environment
list2env(setNames(parameter_diet, names(parameter_diet)), envir = globalenv()) 
# look at items in group 1
g <- 1
# indices of the items in group 1
item_membership_list[g]
\end{minted}
\vspace{-5ex}
\begin{verbatim}
[[1]]
 [1]  1  2  3  4  5  6  7  8  9 10 11
\end{verbatim}

Below we look at the short labels of the items in group 1. For the sake of space, see Supplementary Table S6.1 in \cite{li2023tree} for full description of the items.
\begin{minted}[breaklines,breaksymbolleft=\quad,bgcolor=backgroundCol]{R}
# The name of the list element is the major food group label
item_name_list[g]
\end{minted}
\vspace{-5ex}
\begin{verbatim}
$Dairy
 [1] "dairy_1"  "dairy_2"  "dairy_3"  "dairy_4"  "dairy_5"  
    "dairy_6"  "dairy_7"  "dairy_8"  "dairy_9"  "dairy_10" "dairy_11"
\end{verbatim}

\subsubsection*{Data Simulation}

Data simulation given the true parameter values in ``parameter\_diet" is handled by the ``simulate\_lcm\_given\_tree()" function. Following the dietary assessment example, we simulate a multivariate binary data matrix of $N = 496$ subjects over the $J = 78$ food items, from $K = 6$ latent classes along ``tree\_phylo". The resulting class profiles are weakly separated. Note that the number of latent classes equals the number of leaves in ``tree\_phylo".
\begin{minted}[breaklines,breaksymbolleft=\quad,bgcolor=backgroundCol]{R}
# number of individuals
N <- 496
# random seed to generate node parameters given the tree
seed_parameter = 1
# random seed to generate multivariate binary observations from LCM
seed_response = 1
# simulate data given the parameters
sim_data <- simulate_lcm_given_tree(tree_phylo, N, 
    class_probability, item_membership_list, Sigma_by_group, 
    root_node_location = 0, seed_parameter = 1, seed_response = 1)
\end{minted}

\subsubsection*{Model Fitting}
The primary model fitting function, ``ddtlcm\_fit()", implements a hybrid Metropolis-Hastings-within-Gibbs algorithm to sample from the posterior distribution of model parameters. We assume that the number of latent classes $K = 6$ is known. To use ``ddtlcm\_fit()", we need to specify the number of classes (``K"), a matrix of multivariate binary observations (``data"), a list of item group memberships (``item\_membership\_list"), and the number of posterior samples to collect (``total\_iters"). For a quick illustration, here we specify a small number ``total\_iters = 100".
\begin{minted}[breaklines,breaksymbolleft=\quad,bgcolor=backgroundCol]{R}
set.seed(999)
# number of latent classes, same as number of leaves on the tree
K <- 6
result_diet <- ddtlcm_fit(K = K, data = sim_data$response_matrix, 
  item_membership_list = item_membership_list, total_iters = 100)
print(result_diet)
\end{minted}
\vspace{-5ex}
\begin{verbatim}
---------------------------------------------
DDT-LCM with K = 6 latent classes run on 496 observations and 78 
items in 7 major groups. 100 iterations of posterior samples drawn.
---------------------------------------------
\end{verbatim}

\subsubsection*{Model Summary}

We next summarize the posterior samples using the generic function ``summary()". We discard the first 50 iterations as burn-in's (``burnin = 50"). To deal with identifiability of finite mixture models, we perform post-hoc label switching using the Equivalence Classes Representatives  \citep[ECR,][]{papastamoulis2014HandlingLabel} method by specifying ``relabel = TRUE". To save space in the document, we do not print the summary result (``be\_quiet = TRUE").
\begin{minted}[breaklines,breaksymbolleft=\quad,bgcolor=backgroundCol]{R}
burnin <- 50
summarized_result <- summary(result_diet, burnin, relabel = TRUE, be_quiet = TRUE)
\end{minted}

\subsubsection*{Visualization}
A generic ``plot()" function is available in the package to visualize the summarized result. By specifying ``plot\_option = `all'", we plot both the \textit{maximum a posterior} (MAP) tree and the class profiles. Plotting only the tree or the class profiles is also available through ``plot\_option = `profile'" or ``plot\_option = `tree'".

\begin{figure}
    \centering
    \includegraphics[width=0.9\textwidth]{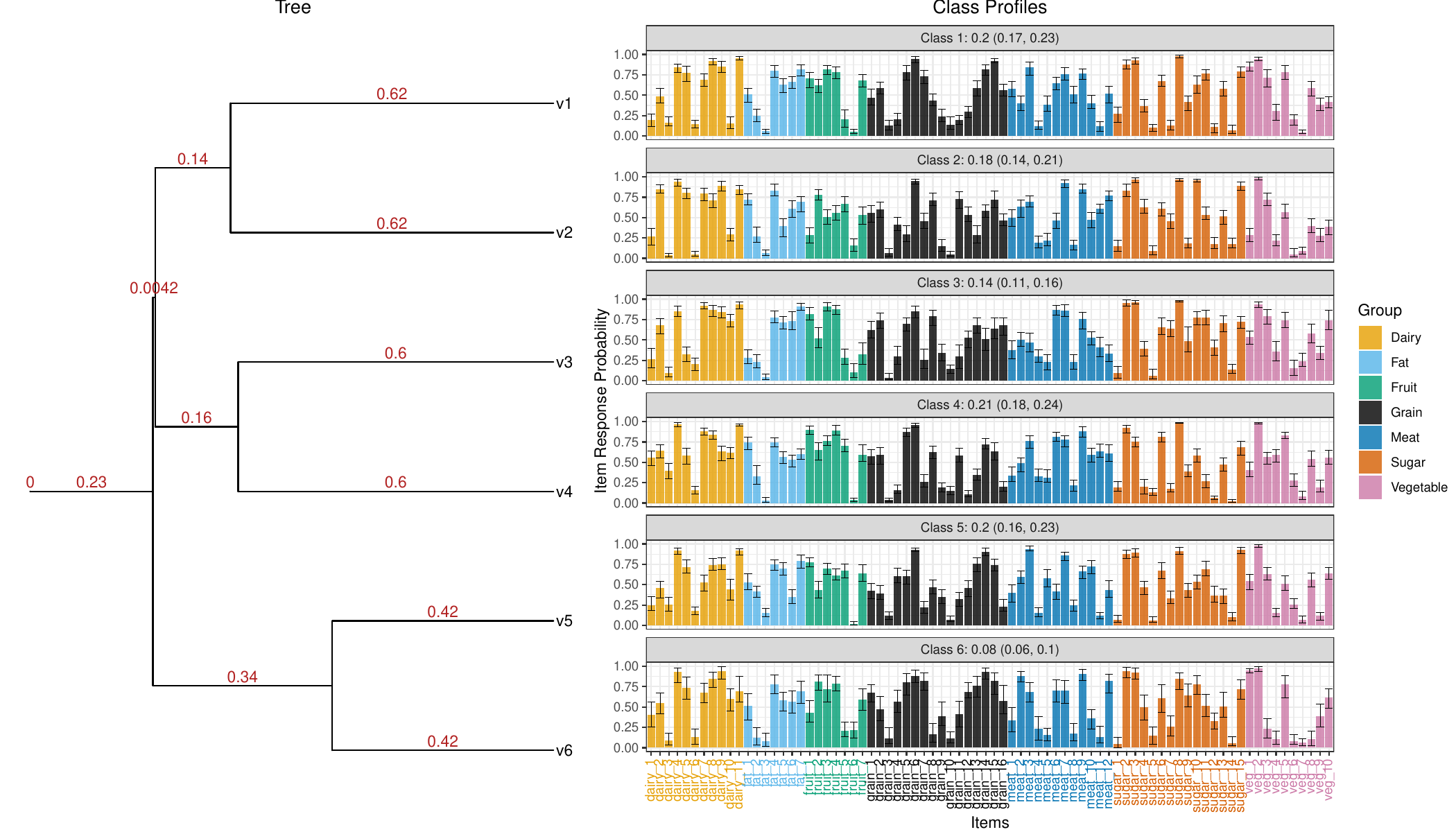}
    \caption{Posterior summary of DDT-LCM with $K=6$ latent classes.}
    \label{fig:summary}
\end{figure}

Figure \ref{fig:summary} displays the result obtained from the above model summary. On the left shows the MAP tree structure over the latent classes. The numbers on the branches indicate the branch lengths. On the right shows class profiles, where major item groups are distinguished by different colors. The description of items is provided in Table S6.1 in the supplement of \cite{li2023tree}. The numbers after the class labels indicate class prevalences along with 95\% credible intervals. Error bars show the 95\% credible intervals of item response probabilities.
\begin{minted}[breaklines,breaksymbolleft=\quad,bgcolor=backgroundCol]{R}
plot(x = summarized_result, item_name_list = item_name_list, plot_option = "all")
\end{minted}

\section*{Interactive Illustration with RShiny}
\sloppy We also provide an accompanying Shiny app with point-and-click interactivity to allow visualization and exploration of model results. The app, available at \href{https://bolinw.shinyapps.io/ddtlcm_app/}{\text{https://bolinw.shinyapps.io/ddtlcm\_app/}}, is designed with three modes, allowing users to 1) simulate data using user-specified parameters or exemplar parameters mimicking a real data set, 2) upload raw multivariate binary observed data, or 3) upload posterior samples collected from a completed fit of the DDT-LCM. Users can explore the app to fully understand the properties of the model, analyze their own data, save the fitted results, and produce visualizations.

\section*{Conclusions}
We developed an R package \texttt{ddtlcm} that addresses a long-standing weak class separation issue in latent class analysis, greatly enhancing numerical and inferential stability relative to existing popular packages. This paper offers a step-by-step example that contextualizes the workflow in a semi-synthetic diet data. Details about usage and more elaborate examples can be found at \href{https://github.com/limengbinggz/ddtlcm}{\text{https://github.com/limengbinggz/ddtlcm}}.

\section*{Acknowledgements}
This work is partially supported by a Michigan Institute for Data Science seed grant. The authors declare no conflicts of interest.

\bibliography{references}
\bibliographystyle{agsm_copy}

\end{document}